\documentclass[useAMS,usenatbib,usegraphicx]{mn2e}
\usepackage{amsmath}
\usepackage[toc,page]{appendix}
\usepackage{pslatex}
\usepackage{amsmath}

\title[Gas kinematics in Perseus spiral arm]{Gas kinematics in massive star-forming regions from the Perseus spiral arm}

\author[]{Kirsanova M. S.$^{1}$, Sobolev A. M.$^2$, Thomasson M.$^{3}$\\
$^{1}$Institute of Astronomy, Russian Academy of Sciences, Moscow, Russia\\
$^{2}$Ural Federal University, Ekaterinburg, Russia\\
$^{3}$Department of Earth and Space Sciences, Chalmers University of Technology, Onsala Space Observatory, Onsala, Sweden\\
}

\begin{document}

\date{kirsanova@inasan.ru}
\pubyear{2017}

\maketitle

\label{firstpage}

\begin{abstract}

We present results of a survey of 14 star-forming regions from the Perseus spiral arm in CS\,(2--1) and $^{13}$CO\,(1--0) lines with the Onsala Space Observatory 20~m telescope. Maps of 10 sources in both lines were obtained. For the remaining sources a map in just one line or a single-point spectrum were obtained. On the basis of newly obtained and published observational data we consider the relation between velocities of the ``quasi-thermal'' CS\,(2--1) line and 6.7~GHz methanol maser line in 24 high-mass star-forming regions in the Perseus arm. We show that, surprisingly, velocity ranges of 6.7~GHz methanol maser emission are predominantly red-shifted with respect to corresponding CS\,(2--1) line velocity ranges in the Perseus arm. We suggest that the predominance of the ``red-shifted masers'' in the Perseus arm could be related to the alignment of gas flows caused by the large-scale motions in the Galaxy. Large-scale galactic shock related to the spiral structure is supposed to affect the local kinematics of the star-forming regions. Part of the Perseus arm, between galactic longitudes from 85$^\circ$ to 124$^\circ$, does not contain blue-shifted masers at all. Radial velocities of the sources are the greatest in this particular part of the arm, so the velocity difference is clearly pronounced. $^{13}$CO\,(1--0) and CS\,(2--1) velocity maps of G183.35-0.58 show gas velocity difference between the center and the periphery of the molecular clump up to 1.2~km\,s$^{-1}$. Similar situation is likely to occur in G85.40-0.00. This can correspond to the case when the large-scale shock wave entrains the outer parts of a molecular clump in motion while the dense central clump is less affected by the shock.

DOI: 
\end{abstract}  

\section{Introduction}\label{Intro}

Methanol masers trace early stages of massive (in the rare cases intermediate) star formation. The most powerful methanol maser transition produces bright emission line at 6.7~GHz, \citet{menten_91}. There are more than 500 sources displaying maser emission at 6.7~GHz in the modern catalogues,~\citet{malyshev, pestalozzi, xu_09, Caswell_MMb}. Methanol masers are associated with very young stages of high-mass star formation (e.g.~\citet{ellingsen}). The statistical analysis performed in \citet{breen_10} reveals a duration of up to 40000 of years of the star formation stage which is accompanied by 6.7~GHz masers. Maser sources at 6.7~GHz belong to the Class~II type. Class~I methanol masers are collisionally excited and are associated with places where molecular outflows interact with the surrounding medium, \citet{sobolev_iau07}. Radiatively pumped Class~II masers are found in less violent regions close to young stellar objects (see observational evidences in recent papers by ~\citet{Hu_16,Chibueze_17} and theoretical investigations by~\citet{Cragg_2005, sobolev_iau05}). It is natural to expect that radial velocities of Class~II masers should be close to the system velocity of molecular gas in regions of star formation. Hopefully it is possible to use the radial velocities of masers for estimation of kinematic distances in statistical studies, \citet{pestalozzi, xu_09, sobolev_iau07}. For example,~\citet{sobolev_iau05} show that there are remarkable peaks on the dependence ``number of maser sources -- distance from the Galactic center''. Locations of these peaks are close to locations of the spiral arms of the Galaxy. A four arm structure of the Galaxy with two bars is favored by study of \citet{green}.

It is still unclear whether maser spots trace dense gas clumps, or regions with coherent velocities,~\citet{sobolev_03}, or both. Anyhow, it is possible to use maser spots for measuring distances by the trigonometric parallax method (see observational review in~\citet{Reid_Honma_2014} and theoretical considerations in~\citet{sobolev_iau12,Sob_2008}), though in the normally avoided cases of the complicated spot structure this method can bring incorrect results, \citet{asaki_14}. This precise method is successfully used in the studies of the spiral structure of the Galaxy when the number of sources is large enough (error is less than 10\% for determination of parameters with 500 sources,~\citet{Honma_15}). Trigonometric parallaxes of a number of methanol maser sources were used by \citet{Reid_2014} for the study of the structure and kinematics of the Galaxy. Their conclusions confirm early results of e.g. \citet{burton_73, georgelin_76, humphreys_76} about non-circular motion of the high-mass star-forming regions in the Perseus arm. Detection of the non-circular motion is in agreement with the theory of galactic spiral density waves. Recent results of \citet{sakai_12, choi_14} show that the methanol maser sources have non-zero velocity component toward the Galactic Center and confirm the existence of the non-circular motion in the Perseus arm.

Previous studies have revealed that the difference between velocity of maser sources and systemic velocity of the dense gas in star-forming regions is not significant, see e.g. \citet{vdW}. They compared velocities of CS\,(2--1) and methanol maser emission at 6.7~GHz for 337 maser features in 63 sources. The distribution of the velocity difference between maser and CS\,(2--1) lines in their Fig.~2 showed almost the same number of red-shifted and blue-shifted masers. Their histogram has a single maximum at 0~km\,s$^{-1}$ and symmetric shape. \citet{malt-45} also compared velocities of dense gas and methanol masers utilizing results of MALT-45 Galactic Plane survey. They made this comparison using CS(1-0) lines and peak velocities of class~I methanol masers at 44.1~GHz. They fitted the distribution of the velocity differences by the Gaussian function with a mean velocity of $0\pm 0.2$~km\,s$^{-1}$. 

All mentioned results were obtained for the samples of methanol masers with coordinates confined to some particular ranges of galactic longitudes and latitudes. The samples are not limited to some source locations in a particular spiral arm of the Galaxy. The mentioned statistical studies considered maser appearance only as a local phenomenon depending on the physical conditions in some specific star-forming region. However, recently possible influence of large-scale phenomena on the structure and kinematics of star-forming regions was reported. For example, authors of \citet{Kretschmer_2013} found remarkable velocity shift of the $^{26}$Al line at 1.8~MeV produced by stellar winds of massive stars and supernovae relative to CO(1-0) emission from \citet{dame} in galactic HI super-bubbles. They explained the shift as a consequence of preferential expansion of the super-bubbles toward the direction of the galactic rotation, see also \citet{Krause_2015}. Their phenomenological model predicts that directions of outflows from young stellar clusters, where $^{26}$Al line arises, should follow the direction of super-bubbles expansion. They also argued that there must be a change in the preferential direction of the outflows from young stellar clusters near the co-rotation radius. The expansion of the super-bubbles in the Galaxy is supposed to be co-directional to the Galactic rotation inside the co-rotation radius and has an inverse direction outside the co-rotation. So, local kinematics in a single star forming region may be controlled by large-scale gas flows in the Galaxy. 

The last example shows that comparison of tracers of various physical conditions in interstellar gas can be useful for linking local and global phenomena. Authors of \citet{sobolev_iau05} analysed the velocity correlation between CS\,(2--1) emission line tracing dense gas and bright 6.7~GHz methanol masers. They found a distinct group of the 6.7~GHz masers with galactic longitudes from 250$^\circ$ to 320$^\circ$ with blue-shifted relative to CS\,(2--1) lines. The group is located in the Scutum-Centaurus spiral arm. Predominance of red-shifted maser sources in the outer part of the Perseus spiral arm was briefly reported by \citet{kirsanova_april06}. Results of \citet{Kretschmer_2013} and \citet{Krause_2015} inspired us to perform further comparisons of gas velocities in maser and thermal regimes.

The aim of this article is a detailed investigation of velocity difference between CS\,(2--1), $^{13}$CO\,(1--0) quasi-thermal lines and methanol maser lines at 6.7~GHz toward high-mass star-forming regions in the Perseus spiral arm. CS\,(2--1) and $^{13}$CO\,(1--0) lines are tracers of dense and more diffuse gas, respectively. We want to explore the effect of galactic spiral shocks on local phenomena in star-forming regions. Radial velocities of masers are used for statistical studies, e.g. to locate spiral arms in the Galaxy, \citet{xu_09}. So, information about non-random differences between the maser velocities and the systematic velocities of the dense gas in star-forming regions can provide important information for such studies.

\section{Source selection and observations}\label{Obs}

We compare velocities of thermal $^{13}$CO\,(1--0) and CS\,(2--1) lines with the maser lines toward 24 out of 26 known 6.7~GHz methanol maser sources in the Perseus spiral arm. The list of  masers is based on \citet{malyshev} catalogue with additions from~\citet{pestalozzi, Xu_2008} and from the 6.7-GHz methanol multibeam maser catalogue of \citet{Green_2012}. The longitude-velocity diagram from~\citet{dame} allows selecting sources from a particular arm. We do not consider sources toward Sagittarius tangent ($l \approx 40-50^{\circ}$) because it is impossible to distinguish between the two spiral arms in this direction. The compiled list is given in Table~\ref{tab:sourcelist}.

\begin{table*}
\caption{List of 6.7~GHz methanol masers discovered the Perseus spiral arm. We observed CS\,(2--1) and $^{13}$CO\,(1--0) emission toward the sources which are marked by bold font. Data about maser positions, distances and near HII regions are taken from 
$^a$\citet{sly99}, $^b$\citet{georgelin_73}, $^c$\citet{cesaroni_99}, $^d$\citet{pest_osoblind}, $^e$\citet{shk00}, $^f$\citet{choi_14}, $^g$\citet{russeil_07}, $^h$\citet{menten_91}, $^i$\citet{reid_09}, $^{j}$\citet{xu_06}, $^k$\citet{lg97}, $^l$\citet{evans_81}, $^m$\citet{Burns_15}, $^n$\citet{snell_90}, $^o$\citet{caswell95}, $^p$\citet{Green_2012}, $^q$\citet{asaki_14}, $^r$\citet{Xu_2008},$^s$\citet{walsh_97},$^t$\citet{caswell_87},
}
\begin{tabular}{|c|c|c|c|c|c|}
  \hline
Source       & RA(2000)   Dec(2000) & $V_{\rm lsr}$: peak ,      & Ref.            & Near HII region       & Sun dist.\\
             & (h:m:s)    (deg:m:s) & interval (km\,s$^{-1}$)    &                 &     & (kpc)    \\
\hline
{\bf G70.14+1.73}  & 20:00:52.6 +33:29:08 & -26.4[-30.0,-22.0] & sly99$^a$       & S99                  & 8.62$^b$ \\
G78.10+3.64  & 20:14:26.0 +41:13:32 & -6.5[-8.0,-4.0]    & sly99           & away from S108       & 1.70$^c$ \\
{\bf G78.62+0.98}  & 20:27:26.8 +40:07:50 & -39.0[-41.0,-38.0] & pe00$^d$        & away from S108       &          -0.4\\
{\bf G85.40-0.00}  & 20:54:13.7 +44:54:08 & -29.5[-33.0,-29.0] & pe00            & S117, VLA G85.4+00.0 &          \\
{\bf G90.91+1.50}  & 21:09:08.1 +50:02:00 & -70.5[-72.0,-68.0] & shk00$^e$       & BG~2107+49           &     \\
{\bf G94.60-1.80}  & 21:39:58.9 +50:14:24 & -43.9[-45.0,-39.0] & sly99           & S124                 & 3.95$^f$ \\
{\bf G98.02+1.44}  & 21:43:01.4 +54:56:16 & -61.6[-62.0,-61.0] & shk00           & VLA G98.04+01.45     &          \\ 
G108.75-0.96 & 22:58:41.3 +58:46:57 & -45.7[-46.0,-45.0] & shk00           & S152                 & 2.39$^g$ \\
{\bf G111.24-0.76} & 23:16:09.3 +59:55:23 & -38.5[-42.0,-37.0] & shk00           & S157                 & 3.34$^f$ \\
G111.54+0.78 & 23:13:45.3 +61:28:11 & -56.0[-62.0,-54.0] & men91$^h$       & NGC7538 (S158)       & 2.65$^i$ \\
{\bf G123.06-6.31} & 00:52:23.7 +56:33:45 & -29.0[-31.0,-27.0] & sly99, shk00    & NGC281-W (S184)      & 2.82$^i$ \\
G133.72+1.22 & 02:25:42.9 +62:06:05 & -44.3[-47.0,-39.0] & sly99           & W3-IRS5              & 1.95$^i$ \\
G133.95+1.06 & 02:27:03.8 +61:52:24 & -44.3[-46.5,-42.0] & men91           & W3(OH)               & 1.95$^j$ \\
G136.84+1.14 & 02:49:23.2 +60:47:01 & -45.4[-46.0,-41.0] & lg97$^k$, sly99 & W5                   &          \\
{\bf G173.48+2.45} & 05:39:12.9 +35:45:54 & -13.6[-16.0,-11.0] & men91, shk00    & S231                 & 1.80$^l$ \\
{\bf G173.59+2.44} & 05:39:27.6 +35:30:58 &      [-14.0,-13.0] & men91           &                      & 1.80$^l$ \\
{\bf G173.69+2.87} & 05:41:37.4 +35:48:49 & -24.1[-25.0,-24.0] & shk00           & S235                 & 1.56$^m$ \\
{\bf G183.35-0.58} & 05:51:10.8 +25:46:14 & -14.5[-16.0,-4.0]  & sly99, shk00    & BFS48                & 2.10$^n$ \\
G188.95+0.89 & 06:08:53.7 +21:38:30 &  11.0[-4.0,12.0]   & men91, cas95    & S247                 & 2.10$^f$ \\
{\bf G189.03+0.78} & 06:08:41.2 +21:31:04 &   9.0[8.0,10.0]    & cas95$^o$      & S247                 &          \\
G189.47-1.22 & 06:02:08.4 +20:09:20 & 18.8[18.5,19.5]	 & gr12$^p$       &                      &          \\
G189.78+0.34 & 06:08:35.5 +20:38:59 &   6.0[2.0,6.0]     & men91, cas95    & S252                 &          \\
G192.60-0.05 & 06:12:54.5 +17:59:20 &   5.0[2.0,6.0]     & men91, cas95    & S255                 & 1.59$^f$ \\
{\bf G196.45-1.68} & 06:14:37.3 +13:49:36 &  15.0[13.0,16.0]   & men91, cas95    & S269                 & 4.05$^q$ \\ 
G212.06-0.74 & 06:47:12.9 +00:26:07 & 48.6[42.7,50.0]    & xu08$^r$       &  at $ \alpha\approx6.5^h, \delta\approx0.5^{\circ}$& \\
G269.46-1.47 & 09:03:14.9 -48:55:11 &  56.0              & wal97$^s$      & GAL269.467-1.48$^t$ & 6.9$^p$  \\
\hline
\end{tabular}
\label{tab:sourcelist}
\end{table*}

We report results of observations toward 14 sources from the list focusing mostly on those without published data on CS\,(2--1) emission. The observed sources are shown with bold in Table~\ref{tab:sourcelist}. Five sources from the list were discovered via survey of 6.7~GHz maser emission toward \textit{IRAS} sources by~\citet{shk00}: G90.91+1.5, G98.02+1.44, G108.75-0.96, G111.24-0.76, G173.69+2.87. Stellar coordinates (J2000.0) for these five maser sources were taken from the \textit{IRAS} point source catalogue, Version 2.0,~\citet{irasps2}. These coordinates are given in Table~\ref{tab:sourcelist}. After the observations had been performed, we have found that two sources out of 26 do not have published data on CS\,(2--1) emission. \citet{bronfman_96} did not find CS\,(2--1) emission toward the southern source G269.46-1.47. The recently discovered G189.47-1.22, \citet{Green_2012}, was not included in our observations. As a result, our list contains 24 out of 26 known 6.7~GHz maser sources which are located in the Perseus arm.

Observations of $^{13}$CO\,(1--0) at 110201.353~MHz and CS\,(2--1) at 97980.968~MHz~\citep{lovas} emission were performed with the 20-m Onsala telescope in 2005 and 2006. Search for $^{13}$CO\,(1--0) and CS\,(2--1) emission and nine-point mapping with 40$''$ step toward selected sources were done in the 2005 session. Large-scale maps of newly observed sources were done during the next session in 2006. Antenna and receiver parameters were the same in 2005 and 2006. The FWHPs of the Onsala telescope beam are 34$''$ and 38$''$ for $^{13}$CO\,(1--0) and CS\,(2--1) lines, respectively. We used a bandwidth of 40~MHz with 1600 channels. Such configuration of the backend corresponded to channel spacing of 0.07~km\,s$^{-1}$. Frequency switching mode was used. Typical system temperature was about 300-500~K for CS\,(2--1) and 600-800~K for $^{13}$CO\,(1--0) line on a $T_{\rm mb}$ scale. Pointing errors were typically 2.3$''$ and 4.2$''$ for azimuth and elevation, respectively. Main beam efficiency was 0.54 for 97~GHz and 0.50 for 110~GHz. So we obtained spectra with a typical noise level of 0.3~K and 0.8~K for CS\,(2--1) and $^{13}$CO\,(1--0), respectively.

Data were reduced with the Onsala XS package\footnote{ftp://yggdrasil.oso.chalmers.se/pub/xs} and 
the CLASS package from the Grenoble GILDAS software\footnote{http://www.iram.fr/IRAMFR/GILDAS},~\citet{petu_05}. We used routine ``GAUSS'' from the package to get parameters of the lines. 

\section{Results}\label{Res}

\subsection{Velocities of thermal and maser lines}

Results of single-point spectrum observations of CS\,(2--1) and $^{13}$CO\,(1--0) lines toward maser positions are shown in Table~\ref{tab:sourceres}. CS velocities taken from the literature are summarised in Table~\ref{tab:litres}.

\begin{table*}
\caption{Parameters of CS\,(2--1) and $^{13}$CO\,(1--0) lines toward the maser sources obtained by us. Formal fit uncertainties are shown in brackets. Values of $V_{\rm CVR} - V_{\rm CSmas}$ used for Fig.~\ref{fig:galview} are shown in the last column.}
\begin{tabular}{|c|c|c|c|c|c|c|c|}
  \hline
Source       & \multicolumn{3}{c}{CS\,(2--1)}              & \multicolumn{3}{c}{$^{13}$CO\,(1--0)} \\
             & $V_{lsr}$    & FWHM      &$T_{\rm mb}$& $V_{lsr}$   & FWHM      &$T_{\rm mb}$& $V_{\rm CVR} - V_{\rm CSmas}$\\
             & (km\,s$^{-1}$)       & (km\,s$^{-1}$)    & (K)       & (km\,s$^{-1}$)       & (km\,s$^{-1}$)    & (K) & (km\,s$^{-1}$)       \\
\hline
 G70.14+1.73 & --20.9 (0.1) & 2.5 (0.2) & 1.0 (0.1) &\multicolumn{3}{c}{not obs.}          &-5.1\\
 G78.62+0.98 & \multicolumn{3}{c}{not obs.}         & --39.0 (0.1) & 2.1 (0.1) &  2.8 (0.1)&\\
             &              &           &           & --47.0 (0.1) & 6.7 (0.1) &  1.0 (0.1)&\\
 G85.40-0.00 & --36.5 (0.1) & 3.8 (0.1) & 1.4 (0.1) & --36.6 (0.1) & 3.9 (0.1) &  8.6 (0.7)& 5.5\\
 G90.91+1.50 & --70.9 (0.1) & 1.3 (0.4) & 0.1 (0.1) & --71.2 (0.1) & 2.9 (0.2) &  0.5 (0.2) & 0.9\\
 G94.60-1.80 & --43.9 (0.1) & 2.7 (0.1) & 2.1 (0.1) & --43.9 (0.1) & 2.4 (0.1) &  8.5 (0.3) & 1.9\\
 G98.02+1.44 & --63.5 (0.1) & 3.1 (0.1) & 2.4 (0.2) & --63.3 (0.2) & 3.7 (0.5) &  1.6 (0.4) & 2.0\\
G111.24-0.76 & --44.5 (0.1) & 3.6 (0.1) & 2.9 (0.2) & --44.2 (0.1) & 2.6 (0.1) & 11.2 (0.4)& 5.0\\
G123.06-6.31 & --30.4 (0.1) & 3.7 (0.2) & 3.8 (0.6) & --30.2 (0.1) & 3.8 (0.1) & 12.1 (0.5)&1.4\\
G173.48+2.45 & --16.4 (0.1) & 4.3 (0.1) & 5.9 (0.4) & --17.0 (0.1) & 3.7 (0.1) & 13.1 (0.6)&2.9\\
G173.59+2.44 & \multicolumn{3}{c}{not obs.}         & --16.7 (0.2) & 3.2 (0.2) &  0.7 (0.3)&\\
G173.69+2.87 & --18.9 (0.1) & 3.5 (0.2) & 0.6 (0.3) & --19.0 (0.1) & 2.0 (0.2) &  4.6 (0.8)&-5.6\\
G183.35-0.58 & --9.3 (0.1)  & 2.5 (0.1) & 7.0 (0.3) &   -9.0 (0.1) & 2.6 (0.1) & 15.2 (0.7)&-0.7\\
G189.03+0.78 &   2.9 (0.1)  & 2.9 (0.1) & 7.1 (0.3) &    2.9 (0.1) & 2.9 (0.1) & 29.9 (1.2)&6.1\\
G196.45-1.68 &  18.5 (0.1)  & 3.7 (0.1) & 1.5 (0.2) &   17.8 (0.1) & 3.7 (0.1) &  9.7 (0.8)&-4.02\\
  \hline
\end{tabular}
\label{tab:sourceres}
\end{table*}

\begin{table}
\caption{Velocities of CS lines toward the maser sources taken from the literature. Values of $V_{\rm CVR} - V_{\rm CSmas}$ used for Fig.~\ref{fig:galview} are shown in the last column. $^a$~\citet{bronfman_96},$^b$~\citet{kim_06},$^c$~\citet{zinchenko_98} $^d$~\citet{wu_11}}
\begin{tabular}{|c|c|c|c|}
\hline
Sources      &  V$_{\rm lsr}$                 & Ref. & $V_{\rm CVR} - V_{\rm CSmas}$\\
             & (km\,s$^{-1}$)                         &     & (km\,s$^{-1}$) \\
\hline
G78.10+3.64  & $V_{\rm CS(2-1)} =-3.8$        & bro96$^a$&-2.2\\
G108.75-0.96 & $V_{\rm CS(2-1)} =-50.3$       & bro96    & 4.8\\
G111.54+0.78 & $V_{\rm CS(2-1)} =-57.2$       & bro96    &-0.7\\
G133.72+1.22 & $V_{\rm CS(2-1)} \approx-38.0$ & kim06$^b$&-5.0\\
G133.95+1.06 & $V_{\rm CS(2-1)} = -46.4$      & bro96    & 2.15\\
G136.84+1.14 & $V_{\rm CS(2-1)} = -39.8$      & bro96    &-3.7\\
G173.59+2.44 & $V_{\rm CS(2-1)} = -16.6$      & zin98$^c$&3.2\\
G188.95+0.89 & $V_{\rm CS(2-1)} = 3.1$        & bro96    &0.9\\
G189.47-1.22 & no data                        &          &\\
G189.78+0.34 & $V_{\rm CS(2-1)} = 8.8$        & bro96    &-4.8\\
G192.60-0.05 & $V_{\rm CS(7-6)} =7.5$         & wu11$^d$ &-3.5\\
G212.06-0.74 & $V_{\rm CS(2-1)} =45.0$        & bro96    &1.35\\
G269.46-1.47 & $V_{\rm CS(2-1)} = -$          & bro96    &\\
\hline
\end{tabular}
\label{tab:litres}
\end{table}

\begin{figure}
\includegraphics[scale=0.7]{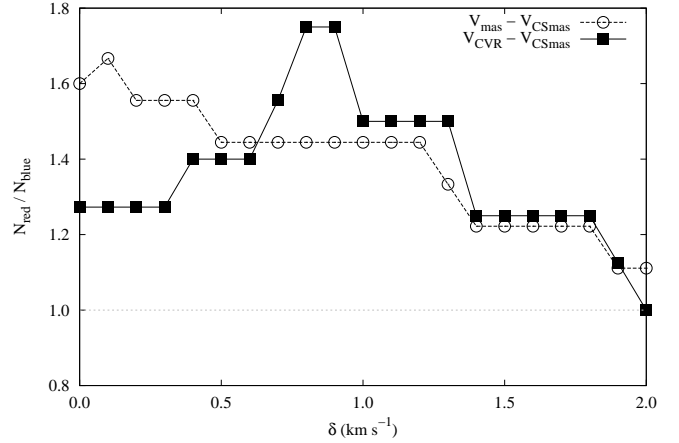}
\caption{Ratio of the red-shifted to the blue-shifted masers for various values $V_{\rm mas}-V_{\rm CS mas}$ (dashed line with empty symbols) and $V_{\rm CVR}-V_{\rm CS mas}$ (solid line with filled symbols)}
\label{fig:nbnr}
\end{figure}

For each source we compare the velocity of the maser peak $V_{\rm mas}$ with the center of the Gaussian fit for the CS line toward the maser position $V_{\rm CS mas}$, and then we do a similar comparison using the center of the maser velocity range $V_{\rm CVR}$ instead of $V_{\rm mas}$. The ratio of the number of sources with $V_{\rm CVR}-V_{\rm CS mas} > \delta$ (we call such sources red-shifted) to the number of sources with $V_{\rm CVR}-V_{\rm CS mas} < - \delta$ (blue-shifted sources) versus $\delta$ is shown in Fig.~\ref{fig:nbnr}. The ratio exceeds 1 for $0<\delta<1.9$~km\,s$^{-1}$. If we consider the similar ratio, using the $V_{\rm mas}-V_{\rm CS mas}$ value, we get a ratio exceeding 1 for $0<\delta<2$~km\,s$^{-1}$. Excess by more than 40\% of red-shifted over blue-shifted masers is found for $V_{\rm CVR}-V_{\rm CS mas}$ and $0.7<\delta<1.4$. We consider the difference between CVR and CS velocity exceeding 0.8~km\,s$^{-1}$ as reliable because it exceeds uncertainties of the velocity determination related to the formal accuracy of the fits, see Table~\ref{tab:sourceres}. It is known that maser emission is variable and the velocity of the maser peak may change. So, comparison of the CVRs with thermal velocities helps confirming previous result obtained for the maser peak velocities. Value of the $\delta \approx 0.7-0.9$~km\,s$^{-1}$ corresponds to the quarter of mean CS\,(2--1) linewidth toward the observed maser sources. An excess of about 20\% of red-shifted on blue-shifted masers remains until $\delta \approx 1.5-1.8$~km\,s$^{-1}$.

\begin{figure}
\includegraphics[scale=0.5]{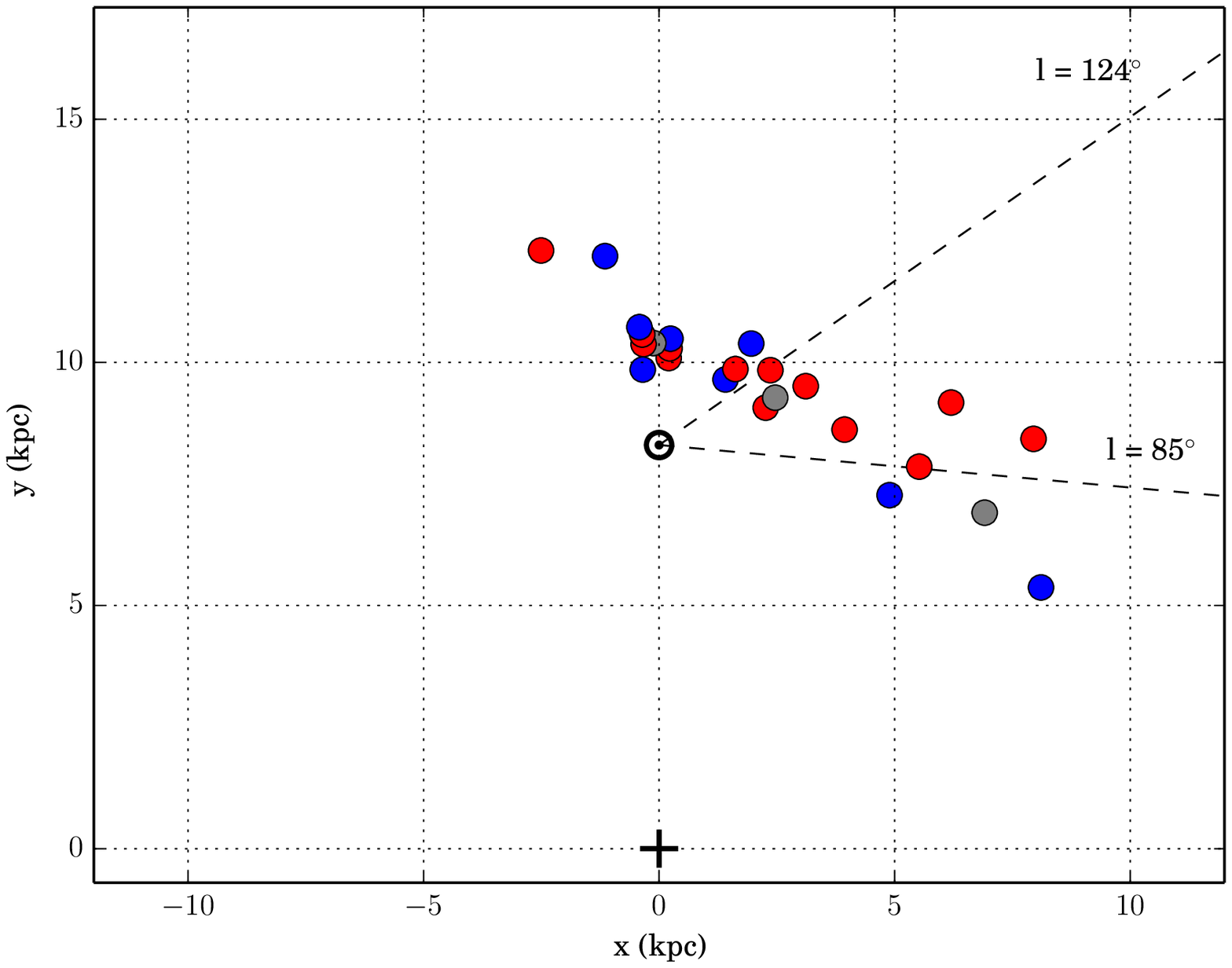}
\caption{Locations of 6.7~GHz methanol masers from the Perseus spiral arm in the Galaxy. Sources with $V_{\rm CVR}-V_{\rm CS mas}<-\delta$, $V_{\rm CVR}-V_{\rm CS mas}>\delta$ and $|V_{\rm CVR}-V_{\rm CS mas}|<\delta$ where $\delta=0.8$~km\,s$^{-1}$ are shown by blue, red and grey respectively. The Galactic center is marked by plus sign at (0, 0) and the Sun
by empty circle with dot at (0, 8.5). Numbers of the red-shifted, blue-shifted and `grey' sources are 14, 3 and 8 respectively. Region with the lack of the blue-shifted masers is outlined by two dashed lines, corresponding galactic longitudes 85$^\circ$ and 124$^\circ$ are shown. This part of the Perseus arm have the greatest radial velocities. Values of $V_{\rm CVR}-V_{\rm CS mas}$ are given in the last columns of Tables~\ref{tab:sourceres} and~\ref{tab:litres}}
\label{fig:galview}
\end{figure}

Locations of the red-shifted and the blue-shifted masers in the Galaxy are shown in Fig.~\ref{fig:galview}. Distances to the sources are taken from the references in Table~\ref{tab:sourcelist}. We use the rotation curve of~\citet{reid_09} to calculate kinematic distances using CS\,(2--1) data to the sources without previously measured distances in the literature. These distances are: $7.05^{+0.63}_{-0.64}$~kpc for G78.62+0.98, $5.54^{+0.64}_{-0.67}$~kpc for G85.40-0.00, $7.95^{+0.70}_{-0.67}$~kpc for G90.91+1.50, $6.62^{+0.67}_{-0.65}$~kpc for G98.02+1.44, $2.86^{+0.72}_{-0.66}$~kpc for G136.84+1.14, $2.16^{+2.93}_{-1.89}$~kpc for G189.78+0.34, $4.72^{+1.22}_{-1.04}$~kpc for G212.06-0.74. Distance to G189.03+0.78 is assumed to be the same as for nearby source G189.78+0.34 because the kinematic method does not provide solution for this source. 

We use $V_{\rm CVR}-V_{\rm CS mas}$ and $\delta=0.8$~km\,s$^{-1}$ for Fig.~\ref{fig:galview}. Red and blue circles show the red-shifted and the blue-shifted masers, respectively. Grey circles correspond to the sources with velocity shift $|V_{\rm CVR}-V_{\rm CS mas}| < 0.8$~km\,s$^{-1}$ which we consider as corresponding to indefinite case. Values of $V_{\rm CVR}-V_{\rm CS mas}$ are given in Tables~\ref{tab:sourceres} and~\ref{tab:litres}. We see almost the same number of red-shifted and blue-shifted masers toward the anti-center of the Galaxy. Radial velocities of the sources toward the anti-center are small, see Table~\ref{tab:sourcelist} and \citet{dame}. A predominance of red-shifted over blue-shifted masers is detected in the second galactic quadrant. 

There is a gap in longitudinal maser distribution between $l \approx 140^{\circ}$ and the {\it anti}-center direction in Fig.~\ref{fig:galview}. The region around  $l \approx 140^{\circ}$ is related with the prominent group of sources from the Bolocam Galactic Plane Survey (BGPS hereafter) \citep{Shirley_2013}, which represent dense molecular regions linked to cluster formation~\citep{Rosolowsky_2010}. Inspection of the BRPS sources locations on the Dame's CO $l-V_{\rm lsr}$ diagram from~\citep{Shirley_2013} reveals that in the Perseus arm they are concentrated between $80^{\circ} < l <  140^{\circ}$. There is also a gap in BGPS sources distribution between $l \approx 140^{\circ}$ and the anti-center direction. So we bring more attention to this part of the Perseus arm and do the statistical study for it. We note again that we do not consider galactic anti-center where projections of peculiar motions on the line of sight become comparable or exceed projections of velocities related to galactic rotation. The number of the red-shifted and blue-shifted sources is 8 and 2, respectively, inside the $80^{\circ} < l <  140^{\circ}$ interval. Binomial distribution gives us $5 \pm 1.6$ red-shifted and blue-shifted sources in the case of random distribution. The confidence of our result that the sources are redshifted and not randomly distributed is thus $1.9\times \sigma$, which corresponds to about 95\%. The part of the Perseus arm between galactic longitudes from 85$^\circ$ to 124$^\circ$ does not contain any blue-shifted masers at all. Sources in this part of the arm have the greatest radial velocities as shown on the longitude-velocity diagram by \citet{dame}.

\begin{figure*}
\includegraphics[scale=0.8]{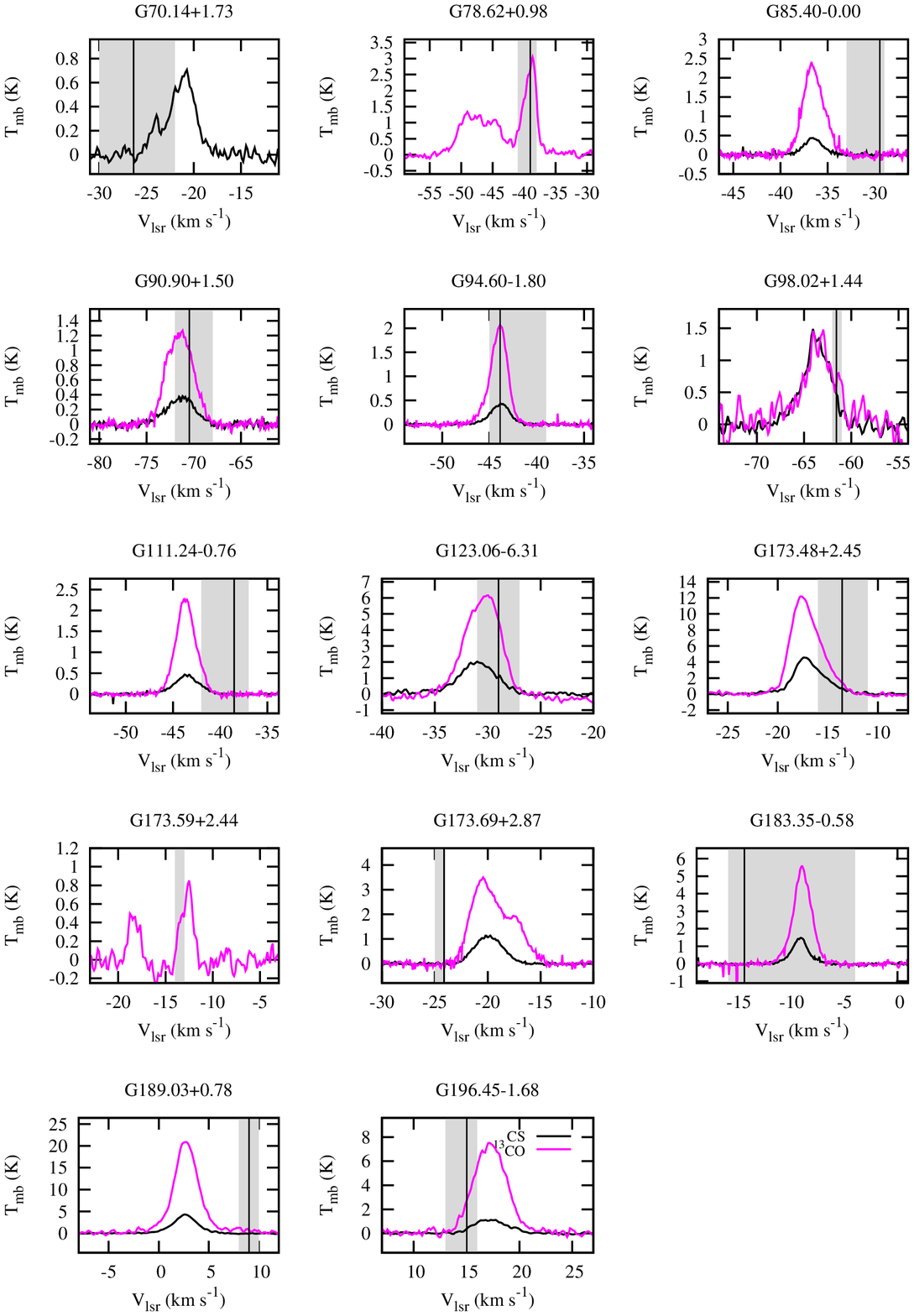}
\caption{Spectra of $^{13}$CO\,(1--0) (magenta) and CS\,(2--1) (black) emission in the observed sources integrated over the whole region. Grey rectangles and vertical lines show the ranges of maser velocities and $V_{\rm mas}$ values according to Table~\ref{tab:sourcelist}.}
\label{fig:aver}
\end{figure*}

We plot the spectra of CS\,(2--1) and $^{13}$CO\,(1--0) emission averaged over all observed positions for each source in Fig.~\ref{fig:aver} to check whether the velocity shifts are inherent just to the maser positions or reflect a general velocity pattern in the star-forming regions. The maser velocity ranges and the maser peak velocities at 6.7~GHz are shown. Velocity ranges of CS\,(2--1) and $^{13}$CO\,(1--0) lines are fully covered by the maser ranges only for 2 out of 14 sources, namely, G78.62+0.98 and G183.35-0.58. Both of them are ``grey''. The maser ranges and the ranges for the thermal lines do not intersect in two red-shifted masers, G85.40-0.00 and G189.03+078. The maser ranges and thermal velocities partially intersect in the other ten sources. Values of $V_{\rm mas}$ and $V_{\rm CVR}$ are shifted relative to the peaks of the averaged spectra. So, methanol maser velocities differ from the velocities of the dense gas in the high-mass star-forming regions of the Perseus spiral arm.

\subsection{$^{13}$CO\,(1--0) and CS\,(2--1) emission maps}

\begin{figure*}
\includegraphics[scale=0.8]{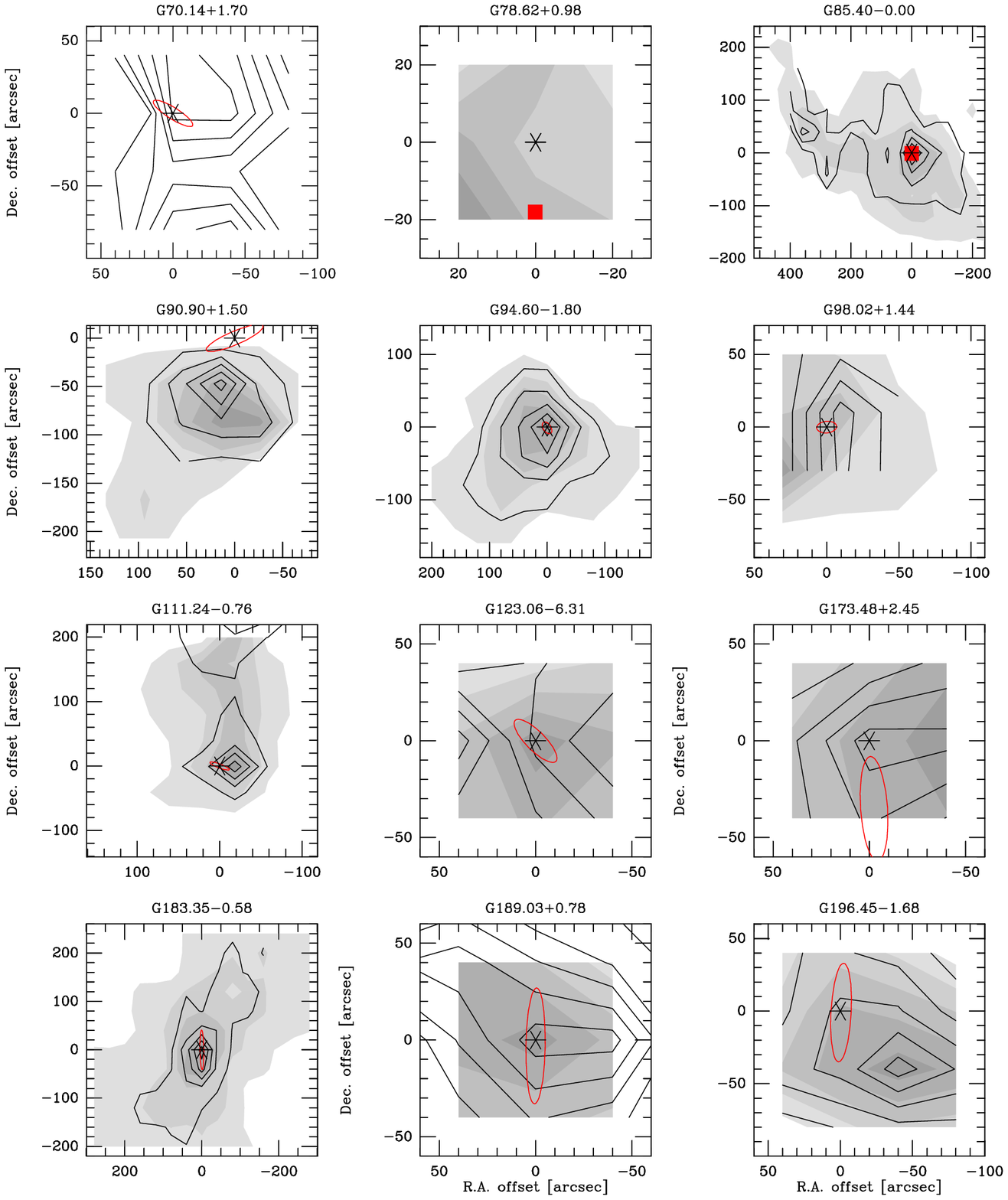}
\caption{Maps of $^{13}$CO\,(1--0) (greyscale) and CS\,(2--1) (contours) emission toward 6.7~GHz methanol masers. \textit{IRAS} and \textit{MSX} sources are designated by ellipses and rectangles respectively. Positions of 6.7~Ghz methanol masers are shown by asterisks. Levels are shown for 10, 30, 50, 70 and 90\% of emission maxima for the observed lines in every source. Zero points for the maps are given in Table~\ref{tab:sourcelist}.}
\label{fig:maps}
\end{figure*}

Maps of the $^{13}$CO\,(1--0) and CS\,(2--1) emission around 6.7~GHz methanol maser sources are shown in Fig.~\ref{fig:maps}. Maps of 10 sources in both lines were obtained. For the rest of the sources a map in one line only or a single-point spectrum were obtained. Large scale maps were made for several sources, namely, G85.40-0.00, G90.90+1.50, G94.60-1.80, G111.24-0.76, G183.35-0.50 and G173.69+2.87. Maps for G173.69+2.87 are published in~\citet{kirsanova_08}. 

Despite the high velocity difference between the thermal and maser ranges in red-shifted sources G85.40-0.00 and G189.03+078, these masers are observed toward the peaks of CS\,(2--1) and $^{13}$CO\,(1--0) emission which also coincide with infrared sources. Same coincidence is observed also in G70.14+1.70, G94.60-1.80, G98.02+1.44, G111.24-0.78 and G183.35-0.58. Masers and infrared sources are both shifted relative to the peaks of the thermal emission in G90.90+1.50, G78.62+0.98 and G196.45-1.68. There is no correlation between relative positions of masers, thermal emission peaks and the sign of the velocity shifts on these maps. We inspected $^{13}$CO\,(1--0) and CS\,(2--1) spectra in the maps of the star-forming regions by eye to find the line wings and check if there is any dependence between appearance of the wings and the sign of the velocity shift. It was found that both red and blue wings are observed in the most of the observed sources not only in the maser position but sometimes in the adjacent area.

\begin{figure*}
\includegraphics[scale=0.55,angle=270]{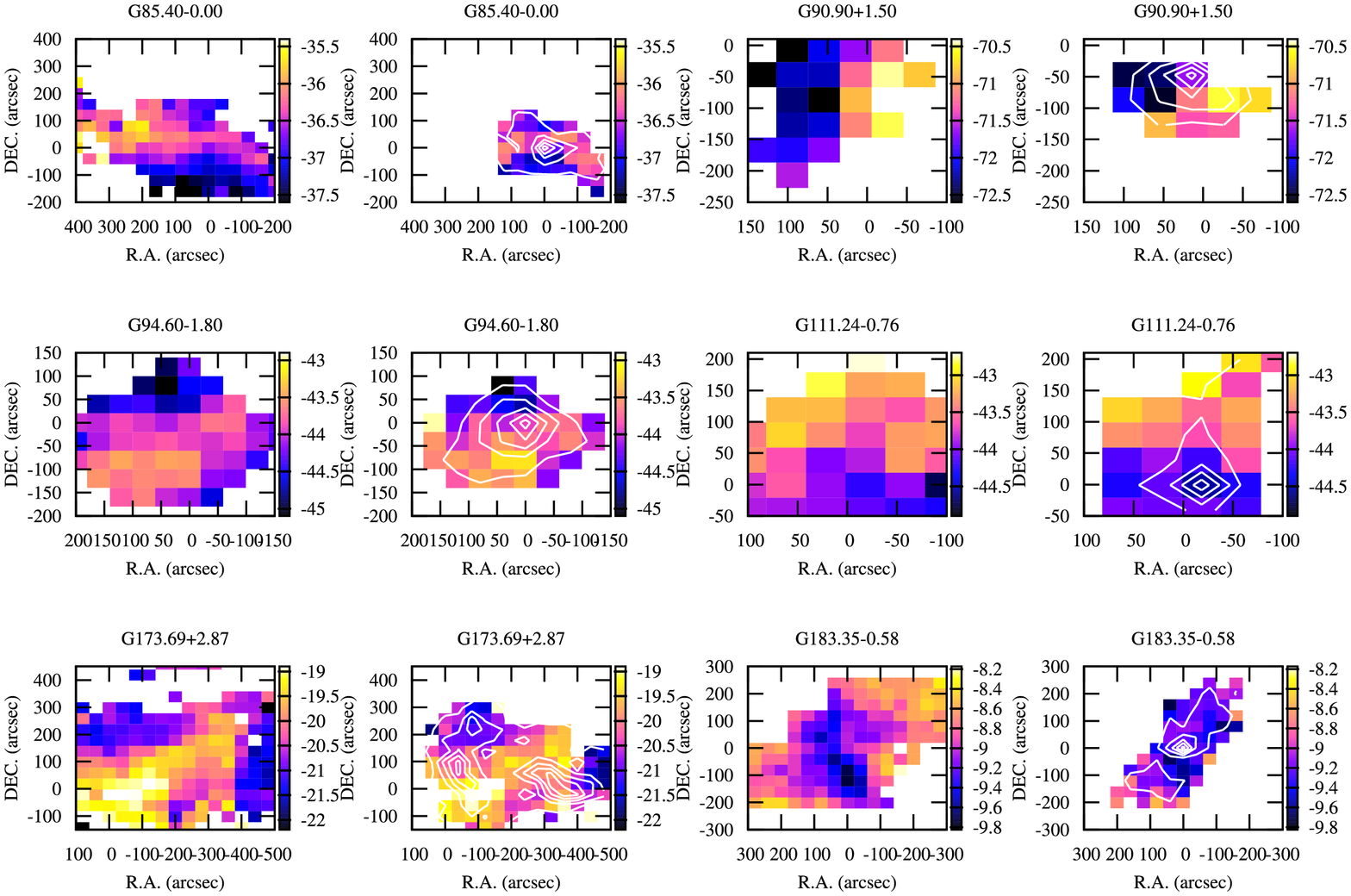}
\caption{Velocity maps of $^{13}$CO\,(1--0) and CS\,(2--1) line emission in km\,s$^{-1}$ are shown by colour. Maps of $^{13}$CO\,(1--0) and CS\,(2--1) emission are shown for each source in the left and right panels, correspondingly. The values of $V_{\rm lsr}$ are taken from the gauss fits of corresponding lines. Size of the colour rectangles corresponds to 40$''$-step of mapping. Maps of CS\,(2--1) integrated emission are shown by white contours on the CS\,(2--1) velocity maps. The contour levels are the same as in Fig.~\ref{fig:maps}. Reference points for the maps are given in Table~\ref{tab:sourcelist}.}
\label{fig:velmaps}
\end{figure*}

To explore the kinematics of the red-shifted and blue-shifted masers in detail we present their velocity maps in Fig.~\ref{fig:velmaps}. After comparison of the integrated intensity maps from Fig.~\ref{fig:maps} and the velocity maps we see that the area of the most negative velocities coincides with the peak of CS\,(2--1) and $^{13}$CO\,(1--0) emission in G183.35-0.58. Velocities at the periphery of the molecular cloud are more positive. Similar velocity distribution is likely present in G85.40-0.00. We note that pronounced difference between velocities of the peak and the periphery in G183.35-0.58 and G85.40-0.00 could be result of the interaction of a molecular cloud with a large-scale wave. Rotation might be responsible for the velocity distribution in G90.90+1.50 and G94.60-1.80. Velocity difference between the dense cores visible in CS\,(2--1) lines is well pronounced in G111.24-0.76 and G173.69+2.87. So, large-scale pattern in the velocity distributions is seen toward one ``grey'' and one red-shifted source.

\begin{figure}
\includegraphics[scale=0.5]{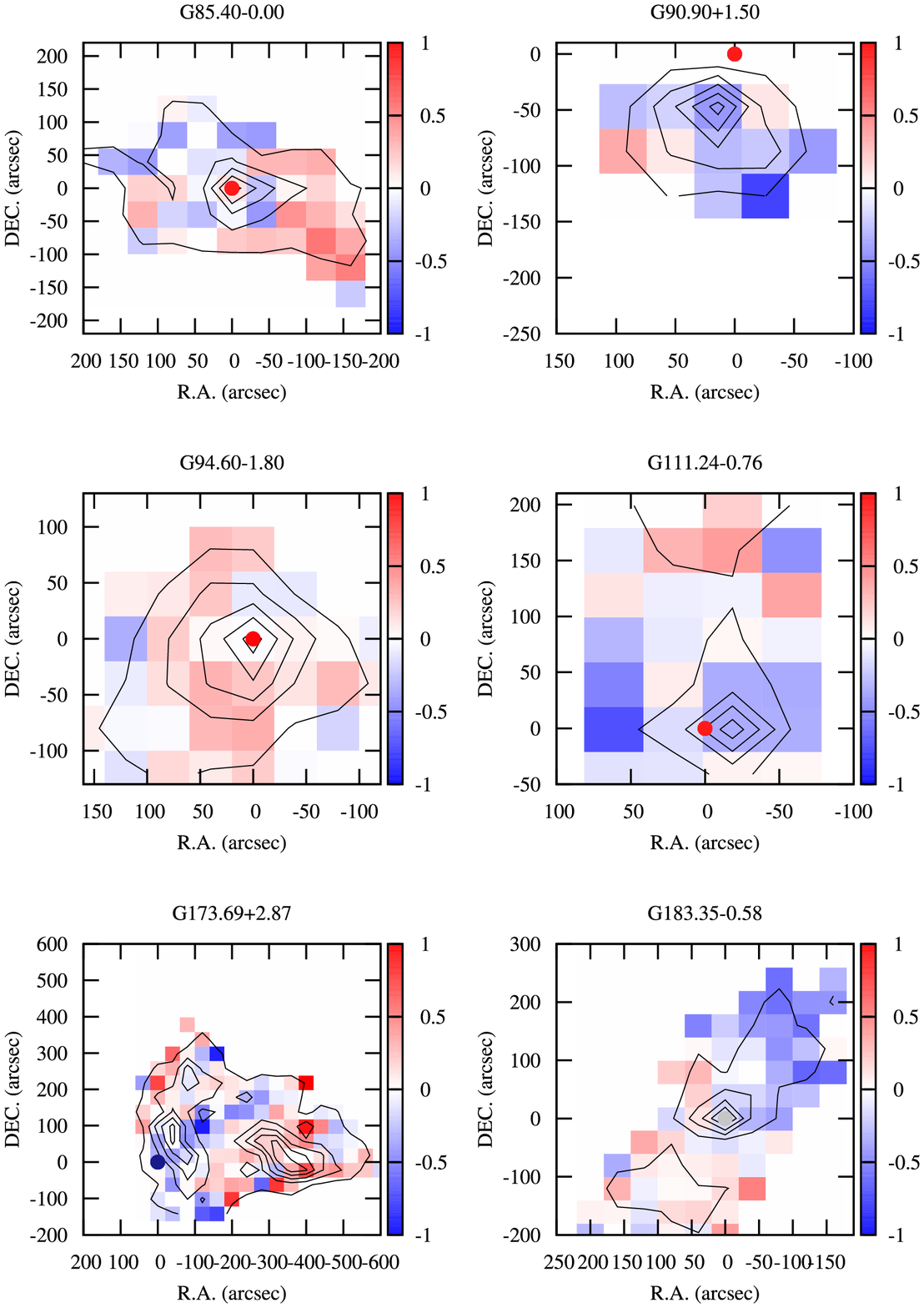}
\caption{Spatial distributions of the velocity differences between CS\,(2--1) and $^{13}$CO\,(1--0) lines: $V_{\rm CS} - V_{\rm ^{13}CO} > 0$ by red and $V_{\rm CS} - V_{\rm ^{13}CO} < 0$ by blue in km\,s$^{-1}$. Size of the rectangles corresponds to the 40$''$-step of mapping. Contours of CS\,(2--1) integrated emission are superimposed on the color maps. The contour levels are same as in Fig.~\ref{fig:maps}. Zero points for the maps are given in Table~\ref{tab:sourcelist}.}
\label{fig:cocsdif}
\end{figure}

We also compare $^{13}$CO\,(1--0) and CS\,(2--1) velocities in the mapped sources by subtracting the value of $^{13}$CO\,(1--0) velocity from the value of CS\,(2--1) velocity in every position of the source. Spatial distributions of the $V_{\rm CS} - V_{\rm ^{13}CO}$ difference are given in Fig.~\ref{fig:cocsdif}. Large-scale area with non-random distribution of the $V_{\rm CS} - V_{\rm ^{13}CO}$ difference is easily seen in the ``grey'' G183.35-0.58 and in G85.40-0.00 with red-shifted masers. The value $V_{\rm CS} - V_{\rm ^{13}CO}$ is positive on one side relative to the CS\,(2--1) emission peak and negative on the other side in these two sources. The value of $V_{\rm CS} - V_{\rm ^{13}CO}$ does not exceed 1~km\,s$^{-1}$ but its spatial distribution is clearly non-random. The value $V_{\rm CS} - V_{\rm ^{13}CO}$ in G94.60-1.80 is positive in almost all positions of the map. Distribution of positive and negative $V_{\rm CS} - V_{\rm ^{13}CO}$ might be not random in G111.24-0.76 also. There is a weak trend from negative values in the south to positive ones in the north but we need larger-scale maps of G111.24-0.76 to prove this. Negative values of $V_{\rm CS} - V_{\rm ^{13}CO}$ are clearly visible toward south-eastern CS\,(2--1) peak in G173.69+2.87 but positive values are observed in the other two. \citet{kirsanova_08,kirsanova_14} related this effect to the internal dynamics of the dense clumps.

\section{Discussion}\label{Disc}

In this article we report the existence of a region in the Perseus spiral arm where velocities of 6.7~GHz methanol masers are predominantly red-shifted relative to the CS\,(2--1) line tracing emission of the bulk of dense gas toward the maser positions. We also found that the total amount of red-shifted masers exceeds that of blue-shifted ones in the entire arm. Currently available data are not sufficient for quantitative interpretation of the result and statistical analysis. Anyhow, we want to give a qualitative astrophysical description of the found velocity shift. 

Methanol masers appear in star-forming regions very close to young stellar objects where dust particles are heated up to 100~K, see~\citet{Sobolev_97, Ostrovskii_2002}. There is a long discussion whether these masers are associated with circumstellar disks (see e.g.~\citet{Norris_98}) or outflow (see e.g.~\citet{DeBuizer_09}). Recent results of \citet{Sanna_15} show that they can be associated with both constituents of the same young stellar source environment. The onset of methanol masers often accompanies development of hyper and ultra-compact HII regions~\citep{sanchez-monge_2011}. These compact objects expand into the surrounding medium preferentially in the direction with the smaller gas pressure,~\citet{Arthur_06,Zhu_15}. It is known that shocks precede ionization fronts during the expansion of HII regions,\citet{Spitzer_1978}. Theoretical calculations and observational data suggest that considerable amount of methanol is released from the grain mantles in locations where the shock front propagates through molecular gas (see e.g.~\citet{Flower_10, Salii_06, Salii_02}). The CS\,(2--1) line is commonly considered as a tracer of a dense gas distribution in star-forming regions. Velocity difference between the methanol masers and CS\,(2--1) emission lines might accompany the expansion of HII regions as well as propagation of shocks and gas outflows in the high-mass star-forming regions. It is not straightforward to predict if methanol masers appear in the directions with lower or higher values of gas number density around expanding ionized regions, but we expect regularity in the velocity difference between thermal and maser lines around them. If methanol masers appear in circumstellar disks we might see greater variety in velocity shifts between masers and dense gas because of the disk rotation,\citet{vdW}.

There is a distinct group of bright blue-shifted sources in the region of the Scutum-Centaurus spiral arm,\citet{sobolev_iau05}. We propose that the existence of regions with non-random difference between velocities of the methanol masers and thermal lines in the Galaxy is caused by processes related to global motions in the Galaxy: galactic rotation, expansion-contraction and spiral arm structure. Our possible scenario implies the existence of a preferential direction for expansion of molecular outflows and HII regions in the Perseus arm. We probably observe a consequence of this preference toward the star-forming regions without significant peculiar motions in the longitude range $85^\circ < l < 124^\circ$. It also brings us to the guess that the red-shifted masers in this range of longitudes are related to expanding HII regions. The blue-shifted masers in the Perseus arm appear toward the galactic anti-center where projections of peculiar motions in star-forming regions on the line-of-sight become comparable or exceed projections of velocities related to galactic rotation. 

Existence of a preferential expansion of outflows and HII regions in the Galaxy was proposed by \citet{Krause_2015}. We have already mentioned their results above. They proposed that since the speed of the spiral pattern is lower than the rotational speed of stars and gas inside the co-rotation radius, stars at the leading edge of an arm preferentially direct their outflows toward galactic rotation into the pre-existing super-bubbles formed by previous generations of stars. Connection between the super-bubbles and CO-emitting gas was also found. \citet{Bagetakos_11} established a spatial correlation between HI shells and super-bubbles created by the activity of massive stars in stellar clusters. \citet{Ehlerova_16} studied spatial correlation between HI shells and CO in the outer parts of the Milky Way. They found an increased occurrence of CO clouds in the walls of the HI shells. So, we expect that expansion of the super-bubbles affects the velocity pattern of CO clouds where star formation occurs. 

It is interesting to consider the mutual location of the co-rotation radius and the location of red-shifted maser sources in the range $85^\circ < l < 124^\circ$ compared with the theoretical description of \citet{Krause_2015}. We have the following arguments to confine location of the masers relative to co-rotation. \citet{loktin_07} studied distribution of open stellar clusters and classical Cepheids in the plane of the Galaxy using wavelet analysis. They found several gaps in the density distribution of younger clusters along the Sagittarius and Local arms filled by the maxima in the density distribution of the older clusters. Older clusters precede younger in the direction of galactic rotation in the second quadrant (see their Fig. 3 and 4). Classical Cepheids show the same tendency. Statistical analysis of \citet{loktin_07} was not done for the objects from the Perseus arm. Anyhow, their results for the objects in the Sagittarius and Local arms are in agreement with the general idea of \citet{Krause_2015}. Comparison of \citet{loktin_07} and \citet{Krause_2015} suggests that the objects from Sagittarius and Local arms are located inside the co-rotation radius.

The model of \citet{Krause_2015} predicts a change of the direction of preferential outflow expansion near the co-rotation radius. Outflows should preferentially expand in the direction of galactic rotation inside the co-rotation radius and in the direction opposite to galactic rotation outside, according to this model. The masers in the second quadrant have negative $V_{\rm lsr}$ relative to galactic rotation but majority of the masers have red-shifted velocities relative to CS\,(2--1). Our scenario suggest that the co-rotation radius is situated outside of the solar circle and the red-shifted masers in the range $85^\circ < l < 124^\circ$. Outflows inside the co-rotation expand toward the galactic rotation in the I and II quadrants of the Galaxy according to \citet{Kretschmer_2013} and \citet{Krause_2015}. Their conclusions point to the existence of the asymmetry between quadrants. Our results are compatible with this asymmetry.

\begin{figure}
\includegraphics[scale=0.9]{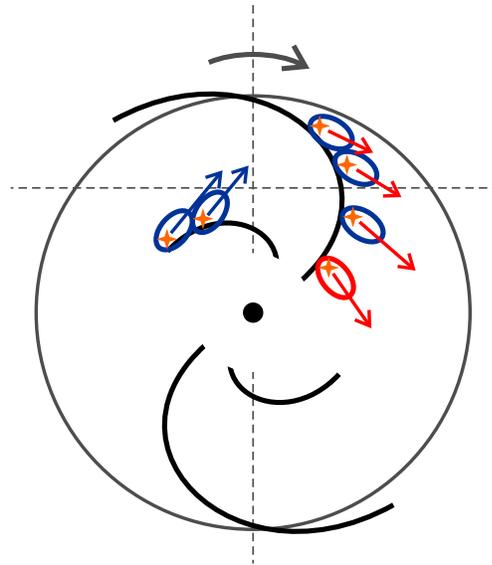}
\caption{Arrangement of the Sun, co-rotation and outflows around young star-forming regions with the methanol masers in our proposed scenario. Direction of galactic rotation is shown by semi-circle arrow. Thin dashed lines divide galaxy plane on quadrants. The Sun is toward the cross of the lines. Schematic spirals are shown for the Perseus and Scutum-Centaurus arms. Co-rotation radius is shown by grey circle. Ellipses on the chart represent super-bubbles created by previous generations of stars. Red and blue color of the ellipses means sign of their $V_{\rm lsr}$. Color arrows show direction of the outflows expansion and sign of their velocity shift. This sign coincides with the velocity shift between the maser sources and thermal gas. Red-shifted maser sources in the range $85^\circ < l < 124^\circ$ with negative $V_{\rm lsr}$ lie inside the co-rotation.}
\label{fig:corotation}
\end{figure}

Fig.~\ref{fig:corotation} shows an arrangement of the Sun, co-rotation and outflows around young star-forming regions with the methanol masers in our proposed scenario. Ellipses on the chart represent super-bubbles created by previous generations of stars. Red and blue colors of the ellipses mean sign of their $V_{\rm lsr}$. Color arrows show the direction of the outflow expansion and sign of their velocity shift. This sign coincides with the velocity shift between the maser sources and thermal gas. So we propose that the red-shifted maser sources in the range $85^\circ < l < 124^\circ$ with negative $V_{\rm lsr}$ lie inside the co-rotation ring where outflows expand toward the direction of galactic rotation. Verification of this hypothesis is out of the scope of this paper, but this issue can be studied in the future.

Parallax-based distances are known for 4 out of 8 sources from the group of masers between $85^\circ < l < 124^\circ$, namely, G94.60-1.80, G111.24-0.76, G111.54-0.78 and G123.06-6.31. We conclude that the co-rotation radius is located outside the solar circle and further than about 10~kpc corresponding to galactocentric distance of G123.06-6.31. This source has the largest galactocentric distance among the red-shifted masers in the range $85^\circ < l < 124^\circ$ with parallax-based distance.

The distinct group of bright blue-shifted sources in the region of the Scutum-Centaurus spiral arm with $l > 310^\circ$ found by ~\citet{sobolev_iau05} is consistent with the proposed scenario. This part of the Galaxy lies inside the co-rotation radius on the chart in Fig.~\ref{fig:corotation} and has $V_{\rm lsr} < 0$. According to \citet{Krause_2015}, outflows and HII regions preferentially expand toward galactic rotation inside the co-rotation, so masers are expected to be blue-shifted relative to the thermal lines.

There are numerous estimates of the co-rotation radius in the literature, and some of them are contradictory. In \citet{Amores_09} and \citet{Dias_05} it was reported that co-rotation radius is located about 10\% further from the Galactic Center than the solar circle. In \citet{russeil_07} deviations from the circular rotation in HII regions from the Perseus and Cygnus arms were studied, and it was found that co-rotation radius is located between them at $\sim 13$~kpc from the Galactic center. \citet{Elias_09} found that spatial distribution and kinematics of stars in Ori\,OB1 and Sco\,OB2 (distance from the Sun is 0.1-0.6~kpc \citep{Melnik_09}) would be explained if they were located close to the co-rotation zone. \citet{Antoja_11} and \citet{Monguio_15} studied stellar kinematics in the solar neighborhood and located the co-rotation radius at about 2~kpc beyond the solar circle. Griv et al. \citet{griv_15} obtained the position of the co-rotation radius about 9.9~kpc.

We note that some other large-scale process(es) could be responsible for the formation of velocity difference between various molecular species. As mentioned before, pronounced difference between velocities of the peak and the periphery in G183.35-0.58 and G85.40-0.00 might be a result of the overflowing of molecular cloud by a large-scale wave. To explore this possibility we need larger-scale maps of CS\,(2--1) and $^{13}$CO\,(1--0) emission which could clarify kinematic features of star-forming regions in the Perseus arm.

\section{Conclusion}

We study velocities of molecular gas in 24 high-mass star-forming regions in the Perseus spiral arm. The sample of the study contains regions with methanol maser emission at 6.7~GHz. The presence of the 6.7~GHz methanol masers is an indication that the process of massive star formation at the early stage is taking place.  We consider 24 out of 26 known 6.7~GHz methanol masers which are clearly associated with the Perseus arm. We do not consider the southern source G269.46-1.47 for which there is no CS\,(2--1) line data in the literature and the recently discovered G189.47-1.22. Our sample does not contain sources in the tangent directions because association of those with the Perseus arm is rather uncertain. Our sample is uniform and consists of high-mass star-forming regions at similar evolutionary stages.

\begin{enumerate}

\item Centers of the radial velocity ranges ($V_{\rm CVR}$ velocity) of 6.7~GHz methanol masers are predominantly red-shifted with respect to the peaks of the thermal CS\,(2--1) lines toward the maser positions ($V_{\rm CS mas}$). We assume that the difference between $V_{\rm CVR}$ and $V_{\rm CS mas}$ velocities exceeding 0.8~km\,s$^{-1}$ is reliable because it exceeds the uncertainties of the velocity determination obtained through the formal fit. The excess of the red-shifted over blue-shifted masers is about 60\%.

\item There is quite a wide region in the Perseus arm, $85^\circ < l < 124^\circ$, where blue-shifted masers with $V_{\rm CVR} - V_{\rm CS mas} < -0.8$~km\,s$^{-1}$ are absent. The number of the red-shifted masers in this region is 7, and one source does not show significant velocity shift. Masers in this particular part of the Perseus arm have the greatest radial velocities. The velocity difference between maser and thermal lines is clearly pronounced here.

\item The existence of the distinct group of the red-shifted masers in the Perseus arm is consistent with the hypothesis about preferential expansion of the outflows in star-forming regions in direction of galactic rotation inside the co-rotation radius.

\item $^{13}$CO\,(1--0) and CS\,(2--1) velocity maps of G183.35-0.58 show difference between gas velocity of the center and the periphery of molecular clump up to 1.2~km\,s$^{-1}$. Similar situation is likely to occur in G85.40-0.00. This can correspond to the case when a large-scale shock wave entrains in motion outer parts of the molecular clump while the dense central clump is less affected by the shock. 

\end{enumerate}

\section{acknowledgements}\label{ackn}
KMS was supported by the President of the Russian Federation grants (MK-2570.2014.2 and NSh-9951.2016.2) and by the OFN program of the Russian Academy of Sciences. AMS was supported by Act 211 Government of the Russian Federation, agreement 02.A03.21.0006. We are thankful to anonymous referee for his/her comments and advises. KMS thanks D.~S.~Wiebe, V.~V.~Krushinsky and N.~N.~Chugai for useful discussions related to this work and to N.~N.~Litvinov for Fig.~\ref{fig:corotation}. The observations were done with the help of L.~E.~B.~Johansson who is deeply missed by us.

SIMBAD database~\citep{simbad}, VizieR~\citep{vizier} catalogue access tool, operated at CDS, Strasbourg, France, NASA's Astrophysics Data System Bibliographic Services were extensively used for the article.

\bibliographystyle{mn2e} 
\bibliography{paper}

\label{lastpage}
\end{document}